\documentclass[usenatbib]{mn2e}
\usepackage{graphicx}
\usepackage{amsmath}
\def\aj{{AJ}}                   
\def\apj{{ApJ}}                 
\def\apjl{{ApJ}}                
\def\aap{{A\&A}}                
\def\mnras{{MNRAS}}             

\def\physrep{{Phys.~Rep.}}   

\title[Conditional Cumulants]{Conditional Cumulants in Weakly Non-linear Regime}
\author[Jun Pan and Istv\'{a}n Szapudi]{Jun Pan\thanks{jpan@ifa.hawaii.edu} 
and Istv\'{a}n Szapudi\thanks{szapudi@ifa.hawaii.edu}
\\Institute for Astronomy, University of Hawaii, 2680 Woodlawn Drive, Honolulu HI 96822, USA}

\newcommand{\avg}[1]{\langle{#1}\rangle}
\newcommand{\mpc}{\rm {h^{-1}Mpc }}

\def\gtsima{$\; \buildrel > \over \sim \;$}
\def\ltsima{$\; \buildrel < \over \sim \;$}
\def\simgt{\lower.5ex\hbox{\gtsima}}
\def\simlt{\lower.5ex\hbox{\ltsima}}

\begin{document}
\maketitle

\begin{abstract}

Conditional cumulants form a set of unique statistics which represent a
sensible compromise between $N$-point correlation
functions and cumulants measured from moments of counts in cells.
They share accurate edge corrected estimators 
with $N$-point correlation functions, 
yet, they are as straightforward to measure and interpret 
as counts in cells. The conditional cumulants have 
three equivalent views as i) degenerate $N$-point
correlation functions  ii) or integrated monopole moments of the bispectrum
iii) they are closely related to neighbour counts.
We compute the predictions of weakly non-linear perturbation theory for
conditional cumulants and compare them with measurements in simulations, both
in real and redshift space. We find excellent agreement between
theory and simulations, especially on scales
$\simgt 20\mpc$. Due to their advantageous statistical properties and
well understood dynamics, we propose conditional cumulants as tools for
high precision cosmology. Potential applications include
constraining bias and redshift distortions from galaxy redshift 
surveys.
\end{abstract}

\begin{keywords}
  Cosmology: theory -- large scale structure of the Universe --
  Methods : statistical
\end{keywords}

\section{Introduction}

Large scale structure statistics of higher than second order
contain a wealth of information on cosmological parameters,
gravitational amplification of initial fluctuations, and
structure formation in general. In particular, higher order statistics
have the potential to provide some of the best constraints on
the phenomenon of ``biasing'' \citep{kaiser84}.
While the core ideas have been known for over two 
decades \citep[cf.][]{Peebles80}, the latest wide
field galaxy surveys, such as the Sloan Digital Sky Survey
\citep[e.g.,][SDSS]{York2000}, and the 2 degree Filed Survey 
\citep[2dF]{Colless2001} have motivated a concerted effort to enhance
theories and techniques of higher order statistics
to the level of ``high precision cosmology''
\cite[for summary, see][]{bcgs2002}.

Counts in cells (CIC) and related statistics have the most
well understood theoretical background and consequently yielded 
some of the most successful applications 
\citep[e.g.,][]{gf94, smn96, szapudietal2002}. 
The estimation methods and the corresponding errors
have been worked out in detail 
\citep[e.g.][]{SzapudiColombi96,scb99}.
But when CIC are measured
to constrain theories at a few percent level,
edge effects present a major problem due to the complex geometry and 
cut out holes of realistic surveys. As shown by \citet{SzapudiColombi96},
edge effects cannot be corrected for exactly; an approximate
estimator exists if the shape dependence of counts is
weak \citep{szapudi98}.

On the other hand, full edge effect correction is possible when
using the class of estimators by \citet{ss98} for
the $N$-point correlation functions. $N$-point correlation functions are,
however, inherently more complicated objects than cumulants of
CIC. They depend on a large number of parameters, their
measurement is computationally intensive \citep{moore2001}, 
and consequently, their interpretation
is difficult. Much of the theoretical effort has been
concentrated on the three-point function, yet
weakly non-linear
perturbation theory 
and halo models have only limited success when contrasted
with simulations \citep{bg2002,tj2003}. 
Redshift distortions present an even
more formidable challenge.

The goal of the present paper is to explore a set of statistics, the
conditional cumulants, 
which combine the simplicity and transparency of cumulants
with the edge effect corrected estimator of the $N$-point correlation.
They can be viewed as degenerate $N$-point correlation functions,
or integrals of the monopole moment of the bispectrum \citep{szapudi2004}.
They are also closely related to (factorial) moments of neighbour counts
\citep{Peebles80}. While previously known, neighbour counts have
been used fairly infrequently in comparison to cumulants 
\citep[e.g.,][]{borgani95}.
The terminology conditional cumulants was introduced 
by \citet{bonometto95}. They have used an estimator based
on moments of neighbour counts and developed a theory under 
the assumptions of stable clustering and scale invariance.

In \S 2, we present formal definition of the conditional cumulants and their
relevant properties. Predictions of the third order conditional cumulant 
in the weakly non-linear regime are described in \S 3. 
In \S 4, we adapt the edge corrected
estimator by \cite{ss98}, and compare predictions
with measurements in simulations. In \S 5, we
summarise results, present the theory in redshift space 
together with simulation results, and discuss
implications for estimating bias.

\section{Conditional Cumulants}

Conditional cumulants are defined as the joint connected 
moment of one unsmoothed and $N-1$ smoothed density fluctuation
fields. They are realised by integrals of the $N$-point
correlation function through $N-1$ spherical top-hat windows,
\begin{equation}
U_N(r_1, \ldots, r_{N-1})= 
  \int \xi_N (s_1, \ldots, s_{N-1},0)
       \prod_{i=1}^{N-1} d^3 s_i \frac{W_{r_i}(s_i)}{V_i} 
\end{equation}
where $V_i = 4\pi/3 r_i^3$ is the volume of the window function 
$W_{r_i}$. In the most general case, each top hat window
might have a different radius.
Further simplification arises if all the top hats are the same, i.e.
we define 
$U_N(r)$ with $r_1=\ldots =r_{N-1}=r$ as the {\em conditional cumulant}
\citep[cf.][]{bonometto95}. 
The $U_N$ subtly differs from the 
usual cumulant of smoothed field $\overline{\xi}_N$ 
by one less integral over the window function.

The second order, $U_2$, is 
equivalent to the confusingly named $J_3$ integral
\citep[e.g.,][]{Peebles80},
\begin{equation}
U_2(r)=\frac{3}{r^3}J_3(r)= \frac{1}{(2\pi)^3}\int P(k) w(kr) 4 \pi k^2 dk\ , 
\end{equation}
where $w(kr)=3(\sin kr-kr \cos kr )/(kr)^3$ is the Fourier transform
of $W_r(s)$, and $P(k)$ is the power spectrum. 

For higher orders, we can construct
reduced conditional cumulants as
\begin{equation}
R_N(r)=\frac{U_N(r)}{U_2^{N-1}(r)}\ .
\end{equation}

$U_N$ and $R_N$ have deep connection with moments
of neighbour counts \citep[e.g.][]{Peebles80}. Let us define
the partition function
$Z[J] = \avg{\exp({\int i J \rho})}$ \citep[cf.][]{ss93},
where $\rho$ is the smoothed density field.
Then we can use the special source function 
$i J(x) = W(x) s+ \delta_D(x) t$ to obtain the
generating function $G(s,t)$. This is related to the
generating function of neighbour counts factorial moments as
$G(s) = \partial_t G(s,t) |_{t = 0}$. The final result is
\begin{equation}
  G(s) =   \sum_{M \ge 1} \frac{(s n V)^M}{M!} U_{M+1}\,
        \exp{\sum_{N \ge 1} \frac{(s n V)^N}{N!} \overline{\xi}_N}\ ,
\end{equation}
where $n V = \bar N$ is the average count of galaxies,
and $\overline{\xi}_1=U_1 =1$ by definition. This generating
function can be used to obtain $U_N$'s and/or $R_N$'s 
from neighbour counts
factorial moments analogously as the generating functions
in \citet{ss93} for obtaining $S_N$'s from
factorial moments of CIC. For completeness,
the generating function for neighbour counts distribution
is obtained by substituting $s \rightarrow s -1$, while
the ordinary moment generating function by $s \rightarrow e^s-1$.
We checked that the we recover formulae
of \citet{Peebles80}, \S 36 from $G(e^s-1)$. 
The above generating
function facilitates the extraction of $U_N$ from neighbour counts
statistics. For details see \cite{ss93}: the entire theory for CIC 
can be adapted to neighbour counts.
So far our discussion has been general;
in what follows we will focus on the first non-trivial
conditional cumulant $U_3$.

$U_3(r_1, r_2)$ is simply related to bispectrum by
\begin{equation}
\begin{aligned}
U_3(r_1, r_2)& =\frac{1}{(2\pi)^6}\int B({\bf k}_1, {\bf k}_2, {\bf k}_3) 
                \delta_D({\bf k}_1+{\bf k}_2+{\bf k}_3) \\
             & w(k_1r_1) w(k_2 r_2)d^3 k_1 d^3 k_2 d^3k_3\ ,
\end{aligned}
\end{equation}
where $\delta_D$ is the Dirac delta function.
To further elucidate the above relation, we use the multi-pole
expansion of the bispectrum and the
three point correlation function proposed by \citep{szapudi2004} 
\begin{equation}
\begin{aligned}
B(k_1, k_2, \theta) &=\sum_{l=0}^\infty \frac{2l+1}{4 \pi} 
B_l(k_1, k_2) P_l(\cos \theta)\ ; \\
\zeta(r_1, r_2, \theta) &= \sum_{l=0}^\infty \frac{2l+1}{4 \pi} 
\zeta_l(r_1, r_2) P_l(\cos \theta)\ ,
\end{aligned} 
\end{equation}
where $\cos \theta$ is ${\bf k}_1 \cdot {\bf k}_2/(k_1 k_2)$ or 
${\bf r}_1 \cdot {\bf r}_2/(r_1 r_2)$, 
and $P_l$ are Legendre polynomials. The multi-pole moments can be obtained 
as 
$B_l=2 \pi \int BP_ld\cos\theta, \ \zeta_l=2 \pi \int \zeta P_l d\cos \theta$.
Substituting into the general equation, we find
\begin{equation}
\begin{aligned}
U_3(&r_1, r_2)=\frac{4 \pi}{V_1 V_2}\int_0^{r_1}\int_0^{r_2} 
\zeta_0(r_1, r_2) r_1^2 r_2^2 dr_1 dr_2\\
=&
\frac{4\pi}{(2\pi)^6}\int dk_1 dk_2 \frac{3k_1}{r_1}
\frac{3k_2}{r_2} j_1(k_1 r_1) j_1(k_2 r_2) B_0(k_1, k_2)\ ,
\end{aligned}
\end{equation}
in which $j_1$ is the first order spherical Bessel function. We see
the $U_3$ depends only on the monopole moment of the bispectrum/three-point
correlation function. This property significantly simplifies the
transformation of the statistics under redshift distortions.

\section{$U_3$ in Weakly Non-linear Regime}

On large scales, where the fluctuations are reasonably small,
clustering of cosmic structures can be understood 
in Eulerian weakly non-linear perturbation theory (EPT) 
\citep[][and references therein]{bcgs2002}.
To predict the behaviour of $U_3$ from Gaussian initial conditions,
we assume an expansion of the density field into first, second, etc.
order,
$\delta = \delta^{(1)}+\delta^{(2)}+\ldots$. Then EPT
can be used to calculate the leading order contribution to
$U_3 = \avg{\delta(0)\delta_r^2}_c$, where $\delta_r$ is the
density field filtered at the scale $r$, and $\avg{}_c$ means
connected moment. We use the second-order EPT kernel \citep{Goroff86}
\begin{equation}
F_2({\bf k}, {\bf k'})=\frac{10}{7}+ \frac{{\bf k} \cdot {\bf k'}}{k k'}
\left( \frac{k}{k'}+\frac{k'}{k} \right) + \frac{4}{7} 
\left( \frac{{\bf k} \cdot {\bf k'}}{k k'} \right)^2\ ,
\end{equation} 
the linear power spectrum $P(k)$, and integrals of the
kernel multiplied with the top-hat window function \citep{b94} to 
finally obtain
\begin{equation}
\begin{aligned}
R_3 (r_1,r_2)\equiv &\frac{U_3(r_1, r_2)}{U_2(r_1) U_2(r_2)}=
\frac{34}{21} \left[ 1+
\frac{\overline{\xi}(r_1, r_2)}{U_2(r_1)} +\frac{\overline{\xi}(r_1, r_2)}{U_2(r_2)} \right] \\
 + \frac{1}{3}&\frac{\overline{\xi}(r_1,r_2)}{U_2(r_1)} \left[ \frac{d \ln U_2(r_2)}{d\ln r_2} +
\frac{\partial \ln \overline{\xi}(r_1,r_2)}{\partial \ln r_2} \right] \\
 + \frac{1}{3}&\frac{\overline{\xi}(r_1,r_2)}{U_2(r_2)} \left[ \frac{d \ln U_2(r_1)}{d\ln r_1} +
\frac{\partial \ln \overline{\xi}(r_1,r_2)}{\partial \ln r_1} \right] \ ,
\end{aligned}
\end{equation}
in which $\overline{\xi}(r_1, r_2)=\frac{1}{2\pi^2}\int k^2 P(k) w(kr_1) w(kr_2)dk$.
The special case when $r_1=r_2=r$ reads
\begin{equation}
R_3=\frac{34}{21}\left[ 1+ 2\frac{\sigma^2}{U_2}\right] +\frac{1}{3}\
\frac{\sigma^2}{U_2}\left[ 2 \frac{d\ln U_2}{d \ln r}+ \frac{d \ln \sigma^2}{d \ln r}
\right]\ ,
\label{eq:r3}
\end{equation}
where $\sigma^2=\frac{1}{2\pi^2}\int k^2P(k)w^2(kr)dk$.
The above equations constitute the main results of this paper.
Note the similarity of $R_3$ with the
skewness, which is calculated in weakly non-linear
perturbation theory as  $S_3 = 34/7- d \ln \sigma^2/d \ln r$
\citep{jbc93, b94b}.

\section{Measurements}

\begin{figure}
\resizebox{\hsize}{!}{
\includegraphics{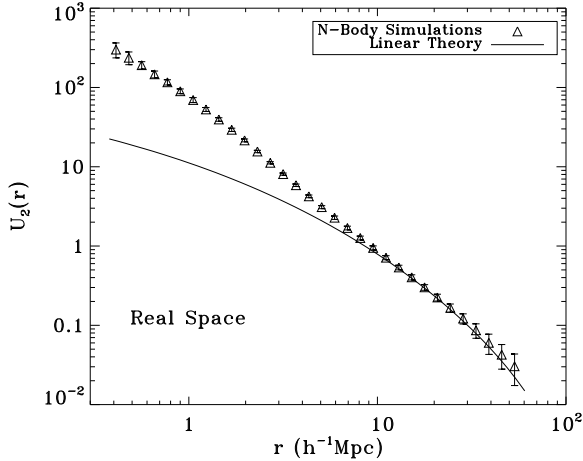}
}
\caption{Predictions of $U_2(r)$ on large scales in real space (solid line)
compared with measurements in N-body simulations
(triangles with error-bars) with $\Lambda$CDM cosmology.}
\end{figure}

\begin{figure*}
\resizebox{\hsize}{!}{
\includegraphics{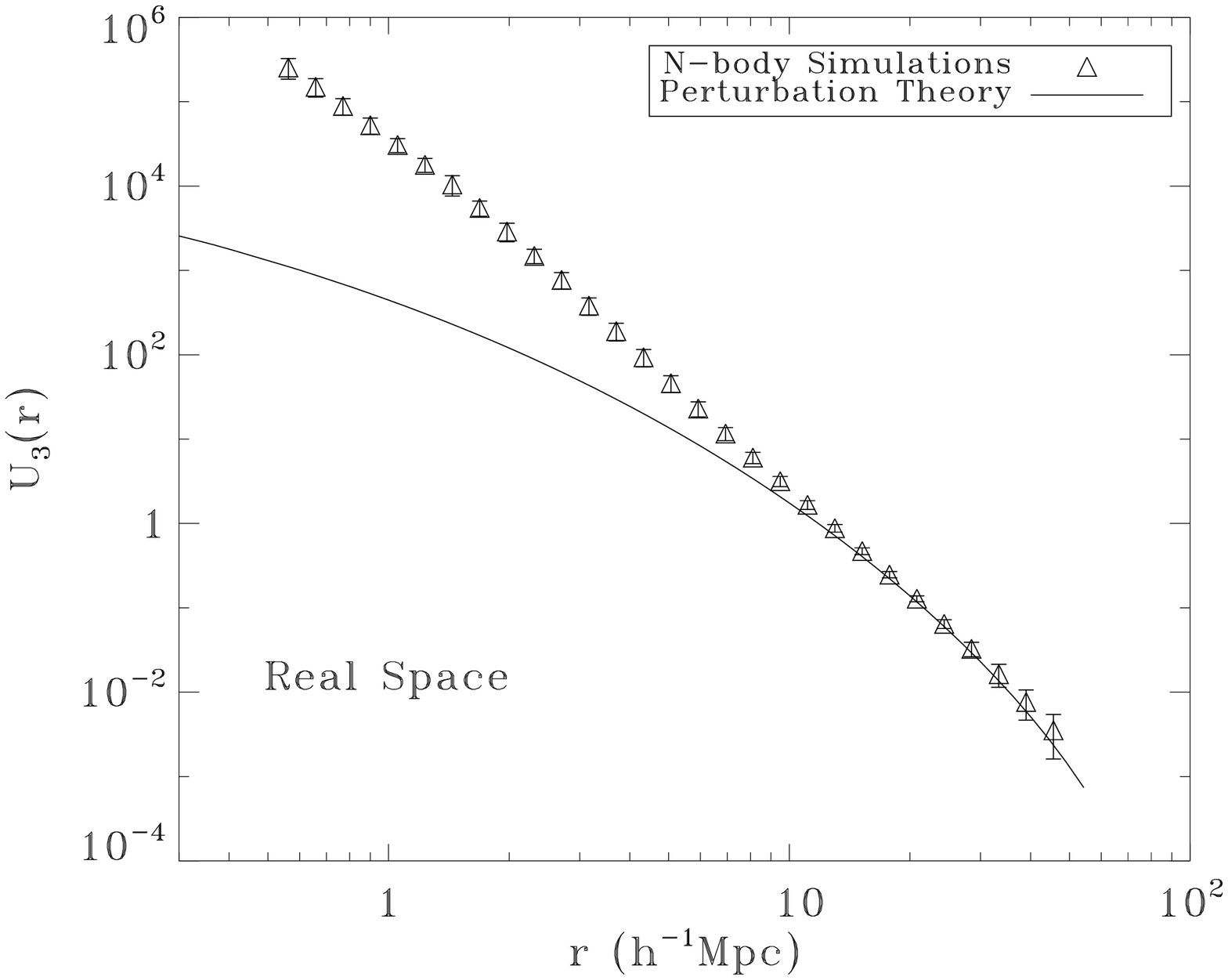}
\includegraphics{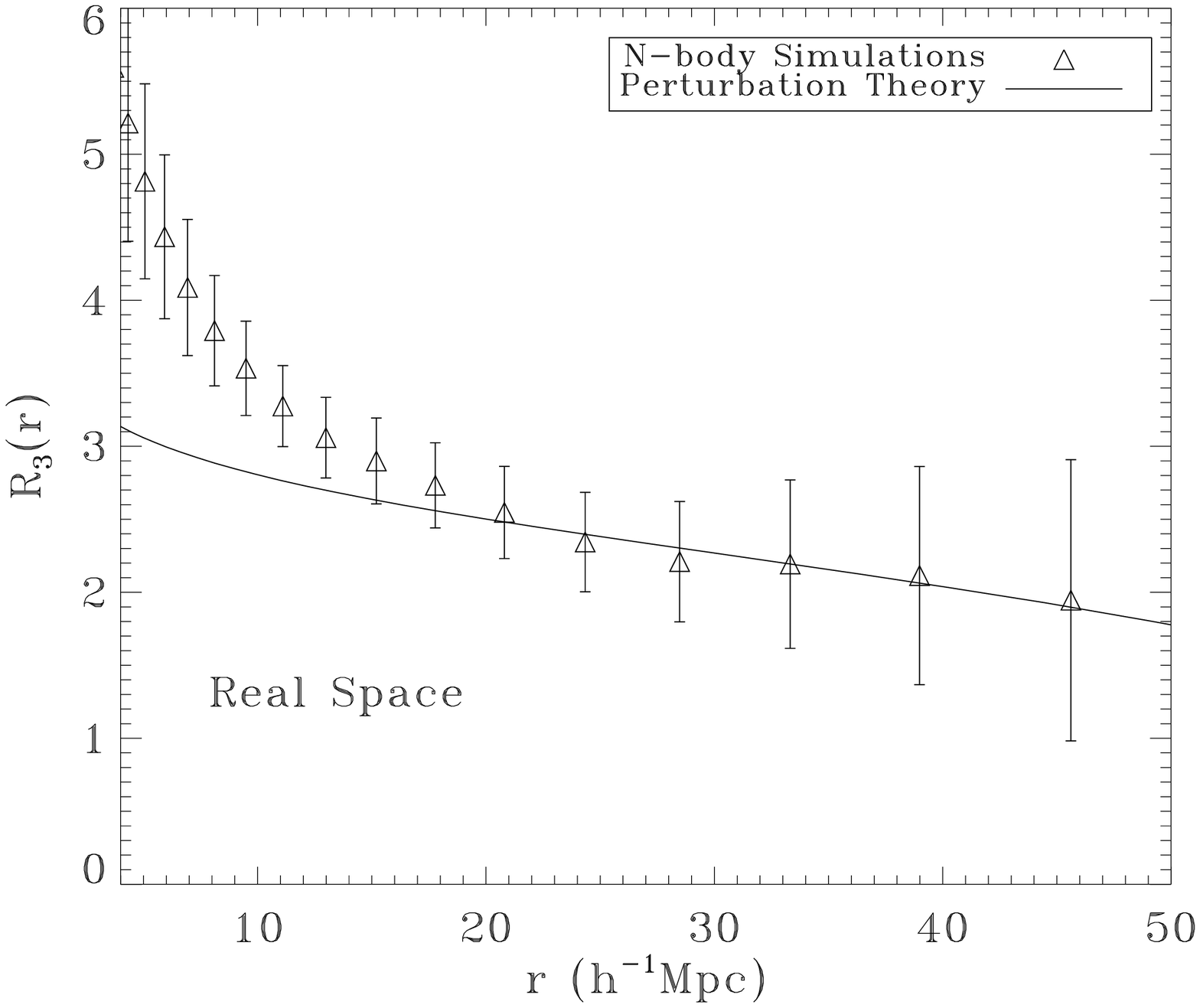}
}
\caption{Same as Figure 1. for the third order conditional cumulant
$U_3$ (left), and the reduced statistics $R_3$ (right).}
\end{figure*}
$U_n(r)$ can be measured similarly to 
$N$-point correlation functions. For instance $U_2$ can be thought
of as a two point correlation function
in a bin $[r_{lo}, r_{hi}] \equiv [0,r]$. Taking
the lower limit to be a very small number instead of 0, one can
correct for discreteness effects due to
self counting (this is equivalent to using factorial moments
when neighbour counts are calculated directly). 
Given a set of data and random points,
the class of estimators of \citet{ss98} will provide an edge (and
incompleteness) corrected technique to measure conditional
cumulants
\begin{equation}
\widehat{U}_n=\frac{(D-R)^n}{R^n}\ .
\label{eq:ss}
\end{equation}
Existing
$N$-point correlation function codes can be used for the estimation;
for higher then third order, one also has to take connected moments
in the usual way.

While the above suggests a scaling similar to $N$-point correlation
functions, the relation to neighbour count factorial
moments outlined in the previous section 
can be used to realise the estimator using two-point
correlation function codes. To develop such an estimator, 
neighbour count factorial
moments need to be collected for each possible combinations
where data and random points play the role of centre and neighbour.

Note that the edge correction of
Eq.~(\ref{eq:ss}) is expected to be 
less accurate for conditional cumulants than for $N$-point
correlation functions, however, the estimator will be
more accurate than CIC estimators. Several alternative
ways for correcting edge effects are known, which
would be directly applicable to conditional cumulants
\citep{Ripley88, Kerscher99, pc2002}. In what
follows, we use  Eq.~(\ref{eq:ss}) for all results presented. 
Future high precision measurements could benefit from
a shootout of possible estimators as in \citet{Kerscher99}.

To test our theory, we performed measurements in $\Lambda$CDM 
simulations by the Virgo Supercomputing Consortium. We 
used outputs of the Virgo simulation and the VLS (Very Large Simulation).
Except for box sizes and number of particles, 
these two simulations have identical 
cosmology parameters: $\Omega_m=0.3$, 
$\Omega_v=0.7$, $\Gamma=0.21$, $h=0.7$ and
$\sigma_8=0.9$. In order to estimate measurement errors, 
we divide the VLS simulation into
eight independent subsets each with the same size and geometry 
the original Virgo simulation. In total, we have used
the resulting nine realisations to estimate errors. 
Note that we corrected for cosmic bias by always
taking the average before ratio statistics were formed.

Our measurements of the second and third order conditional cumulants 
are displayed in Figs. 1 and 2, respectively. Results from EPT (Eq.~\ref{eq:r3})
are denoted with solid lines. The measurements in simulations are
in excellent agreement with EPT, especially on
on large scales $\simgt 20\mpc$.

\section{Summary and Discussion}

We presented the theory of conditional cumulants in
the weakly non-linear regime. The unique set of statistics
can be thought of as degenerate $N$-point correlation
functions, or integrated monopole. We have derived the generating
function of neighbour count factorial moments, revealing a deep
connection to conditional cumulants.
We introduced the reduced quantity $R_N$, which is
analogous to the cumulant $S_N$. We calculated leading order
perturbation theory predictions, and showed that results
are similar to that of the $S_N$'s. However, while edge correction
for CIC is not feasible, we proposed an accurate edge-corrected
estimation method for the conditional cumulants. This 
was applied to a set of measurements in simulations,
which yielded results in excellent agreement with the theory,
especially on  large scales $\simgt 20\mpc$. The agreement with
theory encourages further development of this statistic for
high precision cosmological applications, such as constraining
bias.

As three-dimensional galaxy catalogues are produced inherently in redshift space,
understanding effects of redshift distortions on
these statistics is crucial before practical applications can follow.
In the distant observer approximation, the formula
by \cite{Kaiser87} and \citet{le89} is expected
to provide  an excellent approximation for $U_2(r)$. 
According to \S 2, we only need to consider
the monopole enhancement 
\begin{equation}
U_2(s)=\left( 1+\frac{2}{3}f+\frac{1}{5}f^2 \right) U_2(r)\ ,
\label{eq:pm}
\end{equation}
where $f\approx \Omega_m^{0.6}$. This formula essentially predicts
a uniform shift of the real space results. To test it, we repeated our
measurements in redshift space, and found that the above is indeed
an excellent approximation in redshift space (Fig.~3).

Considering the relatively simple, monopole nature
of the statistics, we expect that the overall effect on $U_3$ should
also be a simple shift, similarly to the Lagrangian calculations by
\citet{hivon95} and the more general Eulerian 
results by \citet{scf99}. 
Specifically, we propose that ratio of
$R_3$ in redshift space to that in real space can be approximated by
\begin{equation}
\frac{ 5(2520 + 3360 f+1260 f^2+9 f^3-14 f^4)}{98(15+10f+3f^2)^2} \cdot 
\frac{7}{4}\ \ \ .
\label{eq:bm}
\end{equation}
This is motivated by the notion that the shift from redshift distortions
of equilateral triangles should be similar to the corresponding
shift for our monopole statistic. Our simulations results (see Fig.~3)
show that this simple idea is indeed a surprisingly good approximation, 
although the phenomenological theory based on the above formula appears to
have $\simeq 5\%$ bias on scales $\simgt 20\mpc$ where
we expect that weakly non-linear perturbation theory is a good
approximation. For practical applications, this bias can be calibrated
by $N$-body, or 2LPT \citep{Scoccimarro2000} simulations.

In addition to the above simple formula, we have calculated the
shift due to redshift distortions 
by angular averaging the bispectrum monopole term in \cite{scf99}.
We have found that the results over-predict redshift distortions,
however, they would agree with simulations at the 1-2\% level
if we halved the terms classified as FOG (finger of god).
At the moment there is no justification for such a fudge
factor, therefore we opt to use the above phenomenology, which
is about 5\% accurate.
While redshift distortions of third order statistics 
are still not fully understood due to the non-perturbative nature of the
redshift space mapping (R. Scoccimarro, private communication),
detailed calculations taking into account velocity dispersion
effects will improve the accuracy of the redshift space theory $U_3$.

\begin{figure*}
\resizebox{\hsize}{!}{
\includegraphics{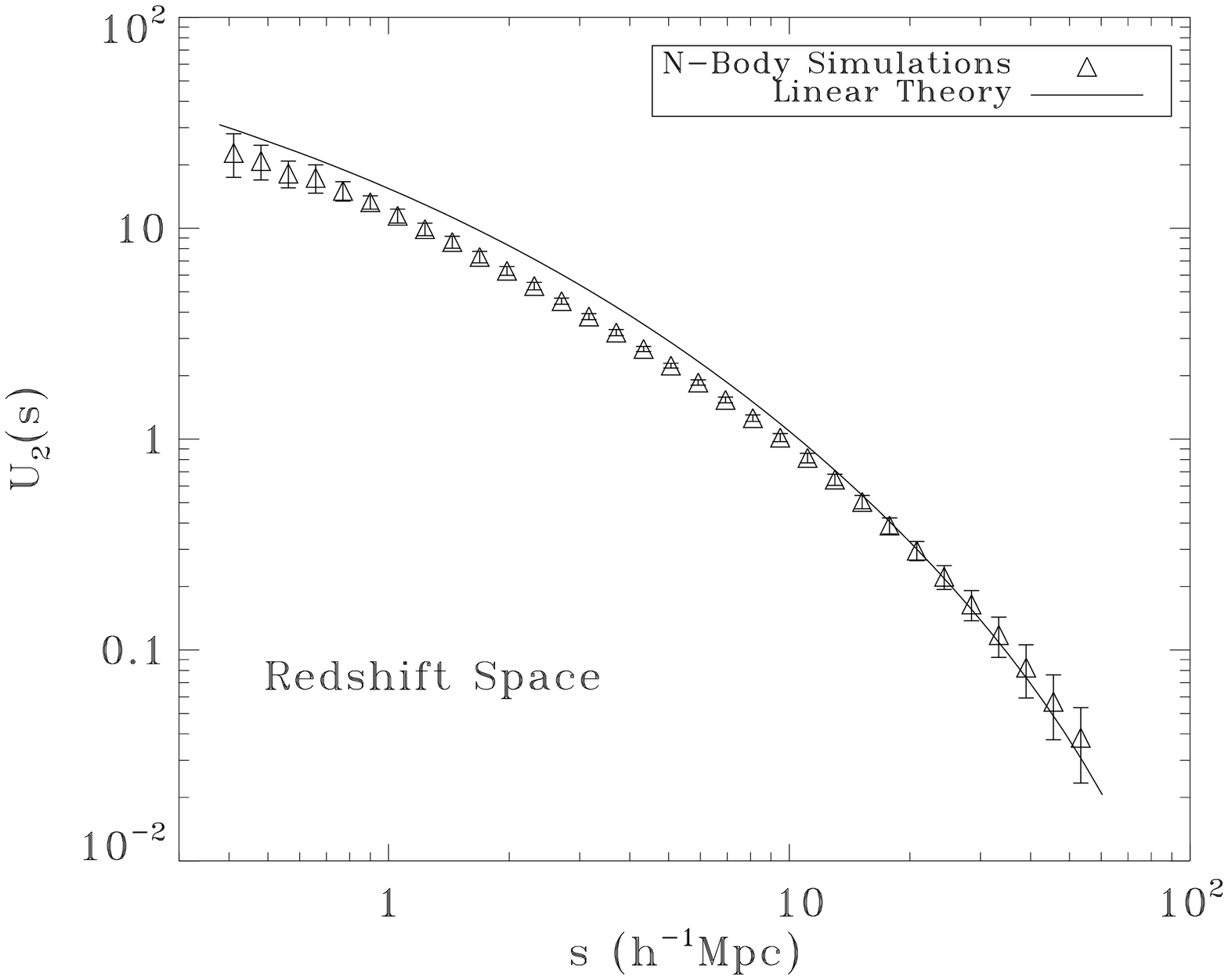}
\includegraphics{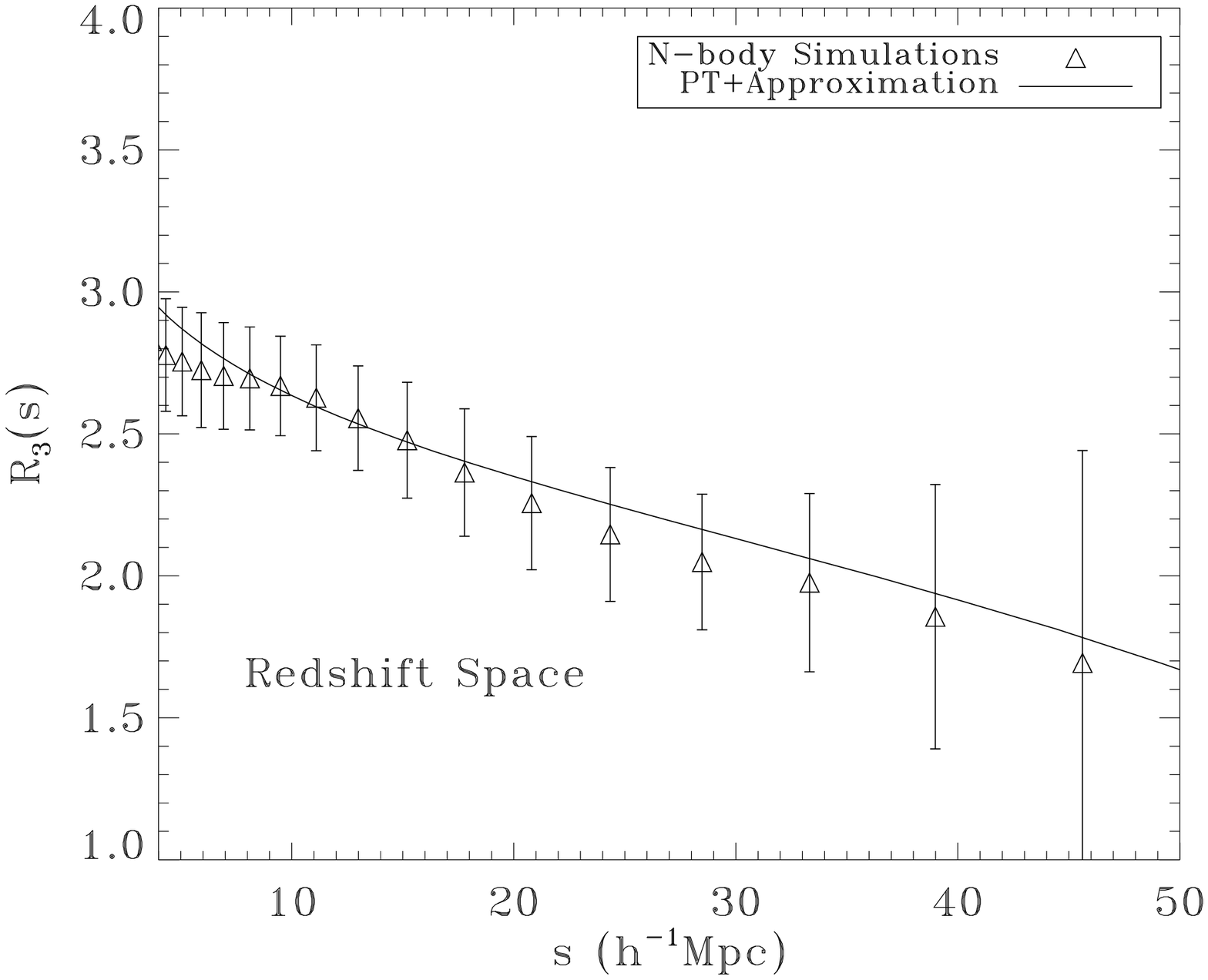}
}
\caption{$U_2$ and $R_3$. The solid line in left panel comes from Eq.~\ref{eq:pm}.
In the right panel the solid line shows
phenomenological model based on Eq.~\ref{eq:bm}, the
theory appears to be a reasonable approximation at the 5\% level.}
\end{figure*}

For applications to constrain bias, one has to keep in mind that redshift
distortions and non-linear bias do not commute. However, at
the level of the above simple theory, it is clear that one can understand
the important effects at least for the third order statistic.
There are several ways to apply conditional cumulants for bias
determination, either with combination with one other statistic
\citep[CIC or cumulant correlators, cf.][]{szapudi98A}, or using
the configuration dependence of the more general $R_3(r_1,r_2)$.
One also has to be careful that in practical
applications ratio statistics will
contain cosmic bias \citep{scb99}. We propose that joint
estimation with $U_2$ and $U_3$ will be more fruitful, even
if $R_3$ is better for plotting purposes. Details of the techniques
to constrain bias from these statistics, as well we determination
of the bias from wide field redshift surveys is left for future work.

Another way to get around redshift distortions is to
adapt conditional cumulants for projected and angular quantities. 
Such calculations are straightforward, 
and entirely analogous to those performed for $S_3$ in the
past.
Another possible generalisation of our theory would be to use
halo models \citep{cs2002} to extend the range
of applicability of the theory well below $20\mpc$. 
These generalisations are left for subsequent research. 

\section*{Acknowledgement}
This work was supported by NASA through grants AISR NAG5-11996,
ATP NAG5-12101, and by NSF through grants AST02-06243 and
ITR 1120201-128440. We thank Pablo Fosalba for stimulating
discussions.
The simulations in this paper were carried out by the Virgo
Supercomputing Consortium using computers based at Computing Centre of
the Max-Planck Society in Garching and at the Edinburgh Parallel
Computing Centre\footnote{The data are publicly available at
{http://www.mpa-garching.mpg.de/Virgo}}.

\bibliographystyle{mn2e}

\end{document}